\documentclass[twocolumn,english,showpacs,floatfix]{revtex4}
\usepackage[T1]{fontenc}
\usepackage[latin1]{inputenc}
\usepackage{amssymb}

\makeatletter

\usepackage{graphicx}
\usepackage{amssymb}

\usepackage{graphicx}

\usepackage{babel}

\usepackage{babel}
\makeatother
\begin{document}

\title{Signature of Mott-insulator transition with ultra-cold fermions in
an one-dimensional optical lattice}

\author{Xia-Ji Liu$^{1}$, Peter D. Drummond$^{1}$ and Hui Hu$^{2,3}$}

\affiliation{$^{1}$\ ARC Centre of Excellence for Quantum-Atom Optics, Department
of Physics, University of Queensland, Brisbane, Queensland 4072, Australia
\\
 $^{2}$\ NEST-INFM and Classe di Scienze, Scuola Normale Superiore,
I-56126 Pisa, Italy \\
 $^{3}$\ Interdisciplinary Center of Theoretical Studies, Chinese
Academy of Science, P. O. Box 2735, Beijing 100084, P. R. China}

\date{\today{}}

\begin{abstract}
Using the exact Bethe ansatz solution of the Hubbard model and Luttinger
liquid theory, we investigate the density profiles and collective
modes of one-dimensional ultra-cold fermions confined in an optical
lattice with a harmonic trapping potential. We determine a generic
phase diagram in terms of a characteristic filling factor and a dimensionless
coupling constant. The collective oscillations of the atomic mass
density, a technique that is commonly used in experiments, provide
a signature of the quantum phase transition from the metallic phase
to the Mott-insulator phase. A detailed experimental implementation
is proposed.
\end{abstract}


\maketitle
\textit{Introduction}. --- The Mott metal-insulator transition (MMIT)
is a fundamental concept in strongly correlated many-body systems.
Recent experiments with ultra-cold atomic gases in optical lattices
are paving the way to explore such quantum phase transitions in a
well-controlled manner. The MMIT with bosonic atoms has been demonstrated
by Greiner \textit{et al.} \cite{greiner}. A demonstration with fermions
has not yet been realized experimentally, although its realization
is within reach of present-day techniques. In the fermionic case,
the relevant theory is the widely studied Hubbard model. This is the
simplest lattice model of interacting fermions, and is exactly soluble
in one dimension.

A gas of single component $^{40}$K atoms in a parallel plane lattice
has already been created by the LENS group, thanks to rapid progress
in the cooling of fermions to temperatures below a micro-Kelvin. This
demonstrated a non-interacting band insulator behavior \cite{lens,hooley,kennedy}
through the observation of dipole mode oscillations. An interacting
gas of ultra-cold fermions with two populated hyperfine levels in
a one-dimensional lattice is also feasible, thus offering the possibility
of observing the fermionic MMIT.

Motivated by this opportunity, we address the problem of how to detect
the emergence of fermionic Mott-insulator phases in real experiments
with ultra-cold fermions. With optical lattices, a harmonic potential
is necessary to prevent the atoms from escaping, so that the Mott-insulator
phase is restricted to an insulator domain at the center of the trap,
and coexists with two compressible metallic wings. Therefore, the
insulator phase cannot be characterized by a global compressibility
as in the unconfined case. Rigol \textit{et al.} \cite{rigol} showed
that a properly defined local compressibility exhibits critical behavior
on approaching the metal-insulator boundary.

In this Letter we show that collective oscillations of the atomic
mass density, an indicator of compressibility, can be utilized to
monitor the emergence of the Mott-insulator phase. We consider a zero
temperature, one-dimensional Hubbard model with a harmonic potential,
as a model of an ultra-cold two-component fermionic atomic cloud in
a deep optical lattice with strong radial and weak axial confinement.
Based on the exact Bethe ansatz solution of the homogeneous 1D Hubbard
model \cite{liebWu}, together with the local density approximation
(LDA), we calculate the density profile of the cloud as functions
of a characteristic filling factor and coupling constant. This leads
to a generic phase diagram including a metallic phase and a Mott-insulator
phase. We then investigate the low-energy collective density oscillations
of the cloud in different phases using Luttinger liquid (LL) theory
\cite{recati}, which describes long wavelength hydrodynamic behaviour.

We find that in the metallic phase the collective oscillation is an
overall motion that goes through all sites of the cloud. This quenches
gradually towards the phase transition point, with the mode frequency
decreasing monotonically to zero. After entering the Mott-insulator
phase, the density oscillation revives, but is restricted to the compressible
wings. Therefore, a sharp dip appears in all collective mode frequencies
in the vicinity of the phase boundary, giving a clear signature of
the MMIT.

\textit{Phase diagram}. --- We consider the 1D Hubbard model \cite{liebWu},
as a prototype for $N$\ ($N/2=N_{\uparrow}=N_{\downarrow}$) interacting
fermions in deep optical lattices, with a Hamiltonian given by: \begin{equation}
{\mathcal{H}}_{0}/t_{h}=u\sum_{i}n_{i\uparrow}n_{i\downarrow}-\sum\limits _{i\sigma}\left(c_{i\sigma}^{+}c_{i+1\sigma}+h.c.\right).\label{LiebWu}\end{equation}
 Here $t_{h}$ and $u$\ $\left(u>0\right)$ are the hopping parameter
and the dimensionless on-site repulsion, respectively. The spin $\sigma=\uparrow,\downarrow$
represents different hyperfine states. The axial harmonic trap adds
a site dependent potential to the Hamiltonian so that:\begin{equation}
{\mathcal{H}}={\mathcal{H}}_{0}+\sum_{i\sigma}\frac{m\omega_{0}^{2}}{2}d^{2}i^{2}n_{i\sigma},\label{model}\end{equation}
 where $\omega_{0}$ is the bare trapping frequency and $d=\lambda/2$
is the lattice periodicity, while $\lambda$ is the wavelength of
standing waves used to create the 1D optical lattice. We describe
the inhomogeneous gas, Eq. (\ref{model}), within the LDA. This amounts
to determining the chemical potential of the gas from the local equilibrium
condition, \begin{equation}
\mu=\mu_{\text{hom}}\left[n\left(x\right),u\right]+\frac{m\omega_{0}^{2}}{2}d^{2}x^{2}.\label{lda}\end{equation}
 We also impose a normalization condition, $\sum_{i=-\infty}^{+\infty}n\left(i\right)=\int_{-\infty}^{+\infty}n(x)dx=N,$
using $\mu_{\text{hom}}\left[n\left(x\right),u\right]$ as the
chemical potential of a homogeneous system with a local filling
factor $n(x)$ ($0\leq n\left(x\right)\leq2$) \cite{coll}. In the
framework of LDA, we replaced the site index $i$ with a
dimensionless variable $x$. The LDA is applicable in the limit of
a large lattice with a slowly varying trapping potential. In other
words, the use of LDA is justified if the size of the gas is much
larger than the harmonic oscillator length
$\sqrt{\hbar/m\omega_{0}}$, implying $\mu\gg\omega_{0}$ or
$N\gg1$. Because of the optical lattice, the single-particle
trapping frequency is renormalized to
$\omega_{eff}=\left(m/m^{*}\right)^{1/2}\omega_{0}$, with an
effective mass $m^{*}=\hbar^{2}/\left(2d^{2}t_{h}\right)$. We also
define a dimensionless frequency,
$\omega=\hbar\omega_{eff}/\left(2t_{h}\right)$, with corresponding
dimensionless time $\tau=t\omega_{eff}/\omega$.

To construct the phase diagram, it is useful to introduce two parameters:

\begin{itemize}
\item $\kappa=u^{2}/\left(16N\omega\right)$ is the interaction strength
\item $\nu=2\sqrt{N\omega}/\pi$ is the characteristic filling factor
\end{itemize}
We clarify the physical meanings of these parameters by considering
two limiting cases.

(i) In the limit of a dilute Fermi cloud {[}$n\left(i\right)\rightarrow0${]},
the Hubbard model reduces to Yang's exactly solvable Fermi gas model.
Therefore, the homogeneous chemical potential takes the form $\mu_{\text{hom}}\left[n,u\right]=\left(2t_{h}\right)\left(\pi^{2}n^{2}/8\right)\mu_{\text{hom}}^{\prime}\left[u/\left(2n\right)\right]$
up to an irrelevant constant. The function $\mu_{\text{hom}}^{\prime}$
is obtained by solving a set of integral equations \cite{yang}.

In terms of the dimensionless chemical potential $\tilde{\mu}=8\mu/\left(u^{2}t_{h}\right)$
and the rescaled coordinate $\tilde{x}=4\omega x/u$, the local equilibrium
and normalization conditions can be re-expressed as: $\pi^{2}/\left[2\gamma^{2}\left(\tilde{x}\right)\right]\mu_{\text{hom}}^{\prime}\left[\gamma\left(\tilde{x}\right)\right]+\tilde{x}^{2}/2=\tilde{\mu}$
and $\int_{-\infty}^{+\infty}2/\gamma\left(\tilde{x}\right)d\tilde{x}=16N\omega/u^{2}$,
where $\gamma\left(\tilde{x}\right)=u/\left[2n\left(\tilde{x}\right)\right]$.
These equations indicate that the coupling strength is $\kappa=u^{2}/\left(16N\omega\right)$,
where $\kappa\ll1$ corresponds to weak couplings and $\kappa\gg1$
is the strongly interacting regime.

(ii) In the non--interacting limit, $\mu_{\text{hom}}\left[n,u=0\right]=-2t_{h}\cos\left(\pi n/2\right)\stackrel{n\rightarrow0}{\approx}2t_{h}\left(-1+\pi^{2}n^{2}/8\right)$.
It is straightforward to show that the central number density is $n\left(x=0\right)=2\sqrt{N\omega}/\pi$,
which we define as the characteristic filling factor $\nu$.

\begin{figure}
\includegraphics[%
  width=9cm]{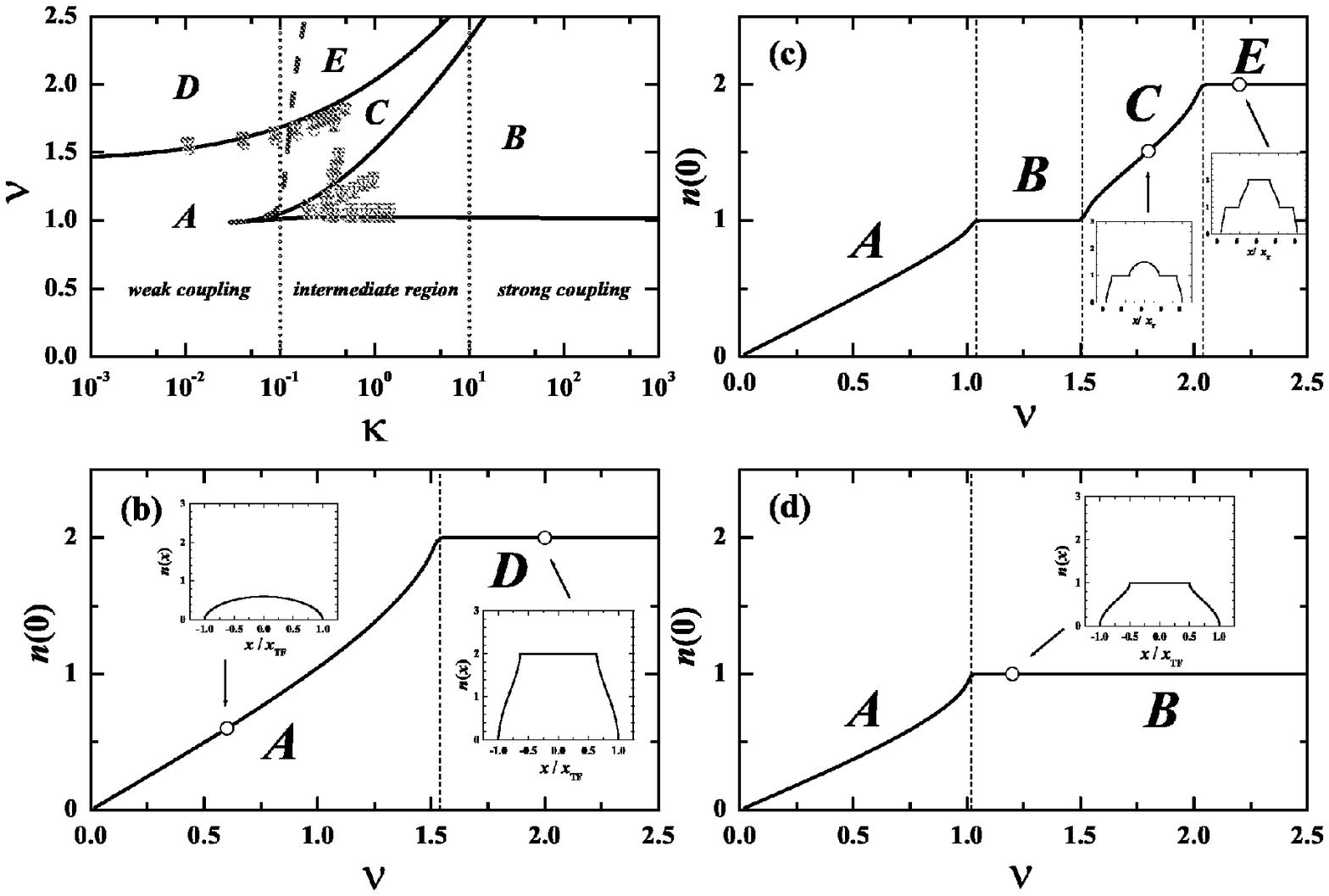}

\caption{(color online) Phase diagram of one-dimensional confined
Fermi gases in optical lattices. $A$: a pure metallic phase. $B$:
a single Mott insulator domain at the center, accompanied by two
metallic wings. $C$:\ a metallic phase at the center surrounded
with Mott insulator plateaus. $D$: a band insulator at the center
with metallic wings outside. $E$: a band insulator at the trap
center surrounded by metallic regions, in turn surrounded by Mott
insulators. We have plotted a dashed line to illustrate the
crossover behavior from $A$ to $C$, or $D$ to $E$. The empty
symbols are the results from quantum Monte-Carlo simulations
\cite{rigol}. In Figs.(b), (c) and (d), the central densities
$n(x=0)$ as a function of $\nu$ have been plotted for three values
of $\kappa=0.01$, $1$, and $100$. The dashed lines in these
figures divide the system into different phases and the
corresponding typical density profiles are shown in insets. }

\label{fig1}
\end{figure}

We determine the phase diagram in Fig.1a from the density profiles.
Five phases (or mixed phases) are identified by plotting the central
density $n(x=0)$ as a function of $\nu$ for different values of
$\kappa$, as shown in Figs. 1b, 1c and 1d, and explained in the figure
caption. In addition, various interaction regimes are illustrated
explicitly by two dotted lines in Fig. 1a, where $\kappa<0.1$ corresponds
to weak couplings, $\kappa>10$, to strong couplings, and an intermediate
regime is in between. In the weak coupling limit ($\kappa=0.01$),
the cloud is noninteracting. An increase of the characteristic filling
factor changes it from a metal to a band insulator. In the strong
coupling regime ($\kappa=100$), the cloud behaves like a noninteracting
\emph{spinless} Tonks gas. It becomes a Mott insulator with increasing
$\nu$, which can also be understood as a band insulator in a spinless
gas. In the intermediate coupling regime, the existence of the trap
leads to a coexistence of metallic and insulating phases, labelled
$C$ and $E$, that have no correspondence in the unconfined case.
An example is shown in Fig. 1c to illustrate the successive emergence
of mixed phases $C$ and $E$ as $\nu$ increases.

A similar phase diagram of ultracold fermions in 1D optical lattices
has been proposed previously in Ref. \cite{rigol} using quantum Monte-Carlo
calculations. Our results for the phase boundaries are in quantitative
agreement with their findings in the same parameter space that has
been simulated. This is expected since the LDA should work well for
these parameters. Since our theory uses an exact analytic solution,
together with the asymptotically exact LDA, this verifies the previous
numerical work. The discrepancy in determining the crossover boundary
from $A$ to $C$ or $D$ to $E$ is due to the use of different criteria.

The phase diagram in Ref. \cite{rigol} covers only the weak coupling
and intermediate regimes of $\kappa\lesssim1.0$. We have extended
their phase diagram to the strongly interacting regime. This extension
turns out to be crucial from the experimental point of view. As we
shall see in Figs. 2a and 2c, the collective modes behave distinctly
in different interaction regimes. Let us now consider in detail these
collective density oscillations which are of primary interest in the
present Letter.

\textit{Collective density oscillations}. --- We describe the low-energy
dynamics of density fluctuations of metallic regions in each phase
using the LL model of the 1D Fermi gas in optical lattices \cite{recati},
where the Hamiltonian can be written as: \begin{equation}
{\mathcal{H}}_{\text{{\textrm{LL}}}}=\sum_{\nu=\rho,\sigma}\int dx\frac{u_{\nu}(x)}{2}\left[K_{\nu}(x)\Pi_{\nu}^{2}+\frac{1}{K_{\nu}(x)}\left(\frac{\partial\phi_{\nu}}{\partial x}\right)^{2}\right].\label{LL}\end{equation}
 Here $u_{\rho}{\textrm{\,\,}}$and $u_{\sigma}$\ are the density
and spin velocities, respectively, and $K_{\rho}$\ and $K_{\sigma}$
are the Luttinger exponents of long-wavelength correlation functions.
These parameters in a spatially homogeneous gas can be calculated
from the Bethe ansatz solution of the 1D Hubbard model \cite{schulz,coll}.
In the Hamiltonian, we have already used the LDA, \emph{i.e.}, $u_{\nu}(x)\equiv u_{\nu}\left(n\left(x\right)\right)$
and $K_{\nu}(x)\equiv K_{\nu}(n\left(x\right))$. The canonical momenta
$\Pi_{\nu}$ are conjugate to the phases $\phi_{\nu}$, \emph{i.e.},
$\left[\phi_{\nu}\left(x\right),\Pi_{\mu}\left(x^{\prime}\right)\right]=i\delta_{\mu\nu}\delta\left(x-x^{\prime}\right)$.
Physically, the gradients of the phases $\partial_{x}\phi_{\nu}\left(x\right)$
are proportional to the density (or spin density) fluctuations, $\delta n_{\nu}\left(x\right)=-\partial_{x}\phi_{\nu}\left(x\right)$,
while the momenta $\Pi_{\nu}\left(x\right)$ are proportional to the
density (or spin density) currents, $j_{\nu}\left(x\right)=u_{\nu}(x)K_{\nu}(x)\Pi_{\nu}\left(x\right)$.

From this linearized Hamiltonian, we derive the hydrodynamic equation
of motion:\begin{equation}
\frac{\partial^{2}\delta n_{\nu}}{\partial\tau^{2}}=\frac{\partial}{\partial x}\left[u_{\nu}(x)K_{\nu}(x)\frac{\partial}{\partial x}\left(\frac{u_{\nu}(x)}{K_{\nu}(x)}\delta n_{\nu}\right)\right].\label{hdyro}\end{equation}
 Hereafter we focus on the readily observable density modes. We consider
the $n$-th eigenmode with $\delta n_{\rho}\left(x\right)\sim\delta n_{\rho}\left(x\right)\exp\left(i\omega_{n}\tau\right)$
and substitute it into Eq. (\ref{hdyro}). The resulting equation
for $\delta n_{\rho}\left(x\right)$ is solved using boundary conditions
which require the current $j_{\rho}\left(x\right)$ to vanish \cite{combescot}.
The details, together with the interesting question of spin-charge
separation, will be presented elsewhere \cite{liu1}.

\begin{figure}
\includegraphics[%
  width=9cm]{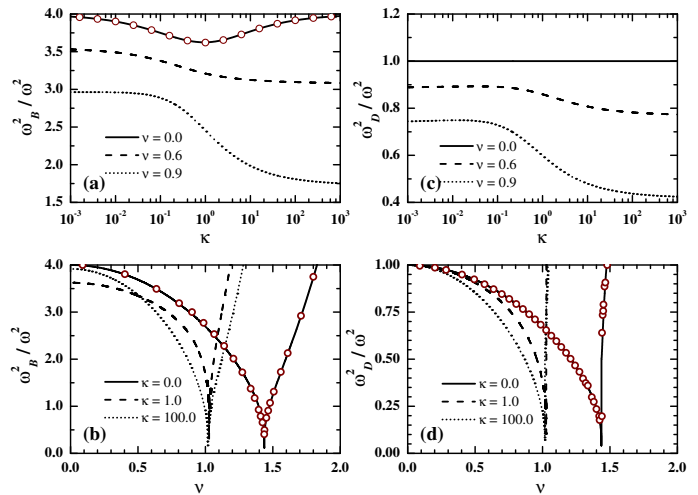}

\caption{Square of frequencies of the breathing mode (Figs. 2a and
2b) and of the dipole mode (Figs. 2c and 2d) at fixed values of
$\nu$ and of $\kappa$ . The circles in (a) are the predictions of
the sum-rule approach, while the circles in (b) and (d) are the
results for the non-interacting Fermi gas, obtained by solving the
semi-classical Boltzmann kinetic equations.}

\label{fig2}
\end{figure}

Fig. 2 shows the square of frequencies of the breathing mode and of
the dipole mode at fixed values of $\nu$ and of $\kappa$. For a
constant value of $\kappa$, the remarkable feature shown in Figs.
2b and 2d is the appearance of a sharp dip at a particular value of
the characteristic filling factor, which has been identified as the
transition point from a metal to an insulator. On the left side of
the transition point, the cloud is a pure metal and the density fluctuates
at all sites. The decrease of the frequency with increasing $\nu$
can be understood from the fact that in the homogeneous case the cloud
has a sound-like spectrum, and the sound velocity $u_{\rho}$ decreases
as $u_{\rho}\propto1-n$ on approaching the transition point at $n=1$,
due to the opening of the charge gap. For larger filling, an insulating
domain forms at the center of the trap, in which the LL description
breaks down. In response to a trap displacement, the center of mass
will develop a quasi-static displacement in this case. Our theory
is able to describe the \emph{local} oscillations of the metallic
wings around this new origin.

We have verified this expectation by using the semi-classical
Boltzmann kinetic theory for a non-interacting cloud
\cite{liu2,kennedy}. As $\nu$ increases, the metallic wings shrink
in size and become less compressible. Therefore the mode frequency
rises after crossing the transition point. This is the central
result of this Letter. The Boltzmann theory also lets us
investigate nonlinear damping effects, which we find increase
rapidly with trap displacement \cite{liu2}. These are absent in
the LL theory, which only treats small displacements.

The application of the LL theory for low-energy dynamics of the 1D
Hubbard model in the unconfined case is already well accepted in
condensed matter physics \cite{schulz}. In the presence of a trap,
we would expect that the LDA is still applicable for describing
the dynamics of the cloud. To this aim, we have examined the
validity of the LL theory within LDA for collective oscillations
by using alternative methods in two limiting cases. (i) In Fig. 2a
the open circles are the prediction of $m_{3}/m_{1}$ or
$m_{1}/m_{-1}$ sum-rules at $\nu\rightarrow0$
\cite{liu1,sumrules}. The excellent agreement shows that the LL
theory is correct at low density for \emph{arbitrary coupling
strengths}. (ii) On the other hand, for a non-interacting gas at a
\emph{finite} filling factor, we calculate the mode frequency by
solving the semi-classical Boltzmann kinetic equation \cite{liu2}.
As shown by open circles in Figs. 2b and 2d, this leads again to
the same result obtained by the LL theory. Combining these two
observations, we expect that the LL theory within the LDA is valid
for finite filling factor at arbitrary couplings, under the
required conditions.

\begin{figure}
\includegraphics[%
  width=7cm]{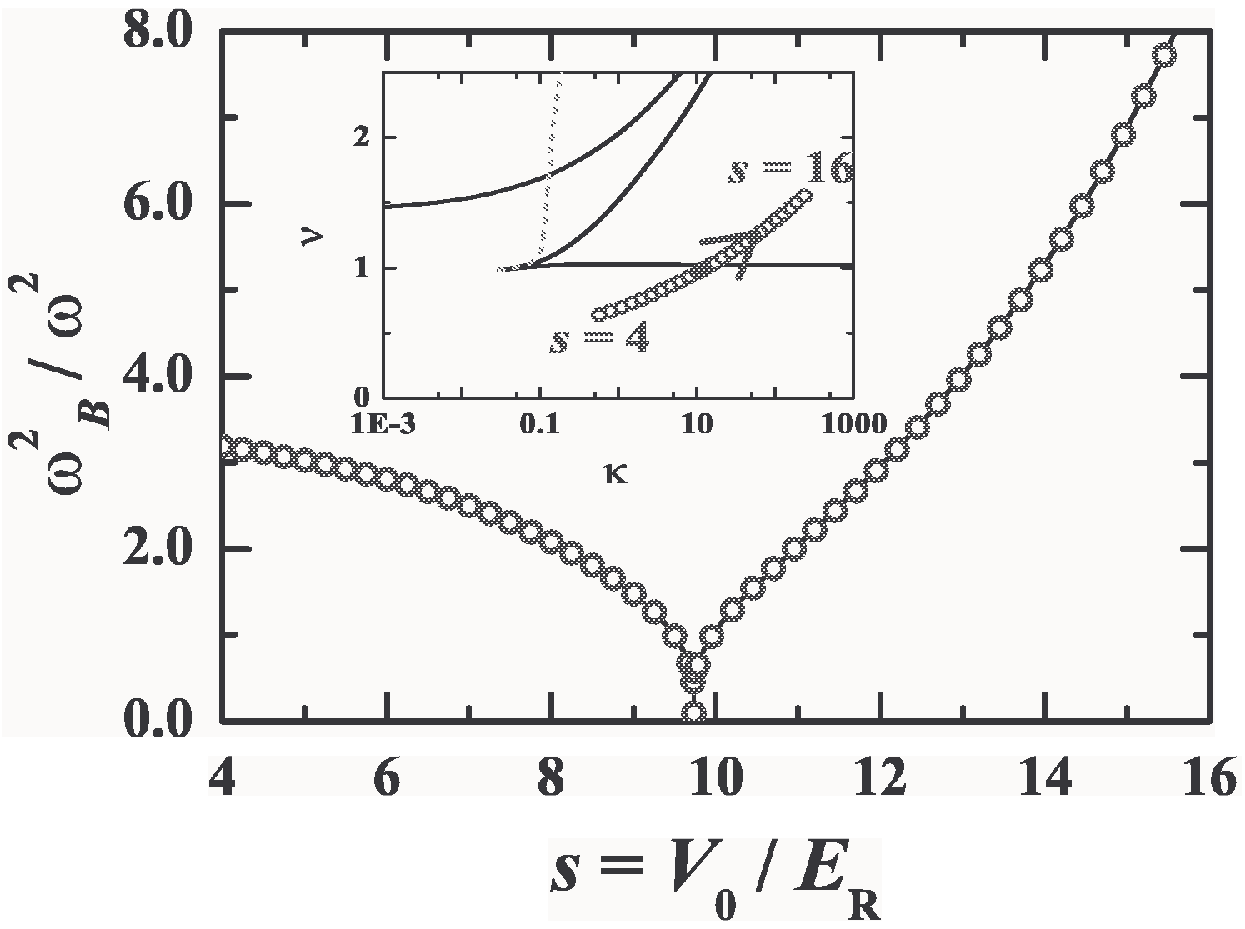}

\caption{(color online) A proposed experimental scheme. The square of breathing
mode frequency has been plotted as a function of the depth of the
lattice potential $s$ for a gas of potassium atoms. The inset shows
the path in the phase diagram with increasing $s$.}

\label{fig3}
\end{figure}

\textit{Experimental proposal.} --- In recent experiments, an
array of quantum gases of bosonic $^{87}$Rb atoms in 1D optical
lattices has been created \cite{paredes}. The gas is first
subjected to a combined potential of a harmonic trap and deep
transverse standing waves. An additional optical lattice with
adjustable depth is then switched on in the axial direction to
form the final configuration. An interacting Fermi gas of $^{40}$K
atoms in 3D lattices has also been demonstrated \cite{kohl}. We
expect that an array of interacting gases of $^{40}$K atoms will
be realized soon. Using the realistic parameters in Ref.
\cite{paredes} (with $^{87}$Rb being replaced by $^{40}$K ), we
have calculated $t_{h}$, $u=U/t_{h}$, and $\omega$ in a Gaussian
approximation \cite{note}. We have taken an inter-species $s$-wave
scattering length $a_{3D}=+26a_{B}$ for $^{40}$K atoms, where
$a_{B}=0.529$ \AA\ is the Bohr radius. The value of $a_{3D}$ is
tunable by a Feshbach resonance. The typical number of atoms in
each tube is $N=15$, which is about the minimum required for the
validity of LDA.

We show the square of frequencies of the breathing mode as a
function of the depth $s$ of the lattice potential in Fig. 3.
Finite temperatures and/or nonlinear effects due to finite
amplitude oscillations, will lead to the decay of collective
modes, and simultaneously weaken the sharpness of the frequency
dip. In addition, the experiments currently average over many
tubes of 1D interacting gases, where the number of atoms has a
parabolic distribution. This will further reduce the dip signal.
Nevertheless, the qualitative picture of the appearance of dip
structure in the vicinity of MMIT transition should persist, and
will be a signature of the entrance into the Mott-insulator phase.

\textit{Conclusions}. --- In summary, we present a phase diagram for
1D confined fermions in an optical lattice. We show that the behavior
of the frequencies of collective density modes is distinct in each
phase. Therefore, the phase diagram can be detected unambiguously
by measuring density oscillations. This provides a useful tool for
locating the quantum phase transition from the metallic phase to the
Mott-insulator phase. We expect that a similar sensitivity of the
mode frequencies with respect to the phase will arise in higher dimensions.
Furthermore, this signature could be applied as well to a gas of 1D
bosons in an optical lattice, since the behavior of its low-energy
excitations in the superfluid phase falls in the same universality
class as the LL theory.

Our investigation is based on the exact Bethe ansatz solution and
the LL theory of 1D Hubbard model, with the effect of harmonic
traps being incorporated in the LDA. We justified the use of LDA
in several limiting cases, both for static and dynamical
properties. To further examine the validity of LDA for dynamics,
more complicated numerical simulations are necessary. Possible
candidates are fermionic phase-space \cite{Corney} or
time-dependent density-matrix renormalization group method
\cite{Kollath}. This is beyond the scope of the present Letter.

We are indebted to M. Rigol and F. Werner for useful discussions.
X.-J. L and P. D. D gratefully acknowledge support by the Australian
Research Council. H. H was partially supported by INFM under the PRA-Photonmatter
Programme.


\begin{thebibliography}{10}
\bibitem{greiner}M. Greiner \textit{et al.}, Nature (London) \textbf{415}, 39 (2002).
\bibitem{lens}L. Pezze \textit{et al.}, Phys. Rev. Lett. \textbf{93}, 120401 (2004).
\bibitem{hooley}C. Hooley and J. Quintanilla, Phys. Rev. Lett. \textbf{93}, 080404
(2004).
\bibitem{kennedy}T. A. B. Kennedy, Phys. Rev. A \textbf{70}, 023603 (2004).
\bibitem{rigol}M. Rigol \textit{et al.}, Phys. Rev. Lett. \textbf{91}, 130403 (2003).
\bibitem{liebWu}E. H. Lieb and F. Y. Wu, Phys. Rev. Lett. \textbf{20}, 1445 (1968).
\bibitem{recati}A. Recati \textit{et al.}, Phys. Rev. Lett. \textbf{90}, 020401 (2003).
\bibitem{coll}C. F. Coll, Phys. Rev. B \textbf{9}, 2150 (1974).
\bibitem{yang}C. N. Yang, Phys. Rev. Lett. \textbf{19}, 1312 (1967).
\bibitem{schulz}H. J. Schulz, Phys. Rev. Lett. \textbf{64}, 2831 (1990).
\bibitem{combescot}R. Combescot and X. Leyronas, Phys. Rev. Lett. \textbf{89} , 190405
(2002).
\bibitem{liu1}X.-J. Liu, P. D. Drummond, and H. Hu, unpublished.
\bibitem{liu2}X.-J. Liu and P. D. Drummond, unpublished.
\bibitem{sumrules}These sum rules are derived by using the formulae in: E. Lipparini
and S. Stringari, Phys. Rep. \textbf{175}, 103 (1989).
\bibitem{paredes}B. Paredes \textit{et al.}, Nature (London) \textbf{429}, 277 (2004).
\bibitem{kohl}M. K\"{o}hl \textit{et al.}, cond-mat/0410389 (2004).
\bibitem{note}The gaussian approximation for the hopping parameter $t_{h}$ becomes
less accurate for a deep lattice depth. A better esimation of $t_{h}$
can be found in: W. Zwerger, J. Opt. B: Quantum Semiclass. Opt., \textbf{5},
S9 (2003).
\bibitem{Corney}J. F. Corney and P. D. Drummond Phys. Rev. Lett. 93, 260401 (2004).
\bibitem{Kollath}C. Kollath \textit{et al.}, cond-mat/0411403 (2004).
\end{thebibliography}
\end{document}